\documentclass[english,aps,manuscript,preprint,showpacs,preprintnumbers]{revtex4}
\usepackage[T1]{fontenc}
\usepackage[latin9]{inputenc}
\usepackage{babel}
\usepackage{amstext}
\usepackage{amssymb}
\usepackage{graphicx}
\usepackage{esint}
\usepackage[unicode=true,
 bookmarks=true,bookmarksnumbered=false,bookmarksopen=false,
 breaklinks=false,pdfborder={0 0 1},backref=false,colorlinks=false,pdfpagemode=FullScreen]
 {hyperref}
\usepackage{breakurl}

\makeatletter
\@ifundefined{textcolor}{}
{%
 \definecolor{BLACK}{gray}{0}
 \definecolor{WHITE}{gray}{1}
 \definecolor{RED}{rgb}{1,0,0}
 \definecolor{GREEN}{rgb}{0,1,0}
 \definecolor{BLUE}{rgb}{0,0,1}
 \definecolor{CYAN}{cmyk}{1,0,0,0}
 \definecolor{MAGENTA}{cmyk}{0,1,0,0}
 \definecolor{YELLOW}{cmyk}{0,0,1,0}
}

\usepackage{amsfonts}
\usepackage{graphics}
\usepackage{epsfig}
\usepackage{bm}

\makeatother

\begin{document}

\title{Complex Scalar DM in a B-L Model }

\author{B. L. Sánchez--Vega}

\email{brucesanchez@anl.gov}

\affiliation{HEP Division, Argonne National Lab, Argonne, Illinois 60439 USA}

\author{J. C. Montero}

\email{montero@ift.unesp.br}

\affiliation{Instituto de Física Teórica--Universidade Estadual Paulista \\
 R. Dr. Bento Teobaldo Ferraz 271, Barra Funda\\
 São Paulo - SP, 01140-070, Brazil }

\author{E. R. Schmitz}

\email{ernany@ift.unesp.br}

\affiliation{Instituto de Física Teórica--Universidade Estadual Paulista \\
 R. Dr. Bento Teobaldo Ferraz 271, Barra Funda\\
 São Paulo - SP, 01140-070, Brazil }
\begin{abstract}
In this work, we implement a complex scalar Dark Matter (DM) candidate
in a $U(1)_{B-L}$ gauge extension of the Standard Model. The model
contains three right handed neutrinos with different quantum numbers
and a rich scalar sector, with extra doublets and singlets. In principle,
these extra scalars can have VEVs ($V_{\Phi}$ and $V_{\phi}$ for
the extra doublets and singlets, respectively) belonging to different
energy scales. In the context of $\zeta\equiv\frac{V_{\Phi}}{V_{\phi}}\ll1$,
which allows to obtain naturally light active neutrino masses and
mixing compatible with neutrino experiments, the DM candidate arises
by imposing a $Z_{2}$ symmetry on a given complex singlet, $\phi_{2}$,
in order to make it stable. After doing a study of the scalar potential
and the gauge sector, we obtain all the DM dominant processes concerning
the relic abundance and direct detection. Then, for a representative
set of parameters, we found that a complex DM with mass around $200$
GeV, for example, is compatible with the current experimental constraints
without resorting to resonances. However, additional compatible solutions
with heavier masses can be found in vicinities of resonances. Finally,
we address the issue of having a light CP-odd scalar in the model
showing that it is safe concerning the Higgs and the $Z_{\mu}$ boson
invisible decay widths, and also the energy loss in stars astrophysical
constraints.

\end{abstract}
\maketitle

\section{Introduction}

\label{sec:introduction}Currently, it is well established from several
observations and studies of the Universe on different scales that
most of its mass is constituted of dark matter (DM) ~\cite{blumenthal1984,davis1985,clowe2006,bennettWMAP,planck2013}.
Although, the nature of DM is still a challenging question, the solution
based on the existence of a new kind of neutral, stable and weakly
interacting massive particles (WIMPs) is both well motivated and extensively
studied. This is mainly due to two reasons. The first reason is that
WIMPs appearing in a plethora of models ~\cite{goldberg1983,ellis1984,kane1994,edsjo1997,hooper2007,agashe2004,birkedal2006,burgess2001,barger2009,okada2010,lindner2011}
give ``naturally'' the observed relic abundance, $\Omega_{\textrm{DM}}h^{2}=0.1199\pm0.0027$~\cite{planck2013}.
The second reason is that WIMPs may be accessible to direct detection.
Currently, there is a variety of experiments involved in the search
for direct signals of WIMPs which have imposed bounds on spin-independent
WIMP-nucleon elastic scattering~\cite{XENON1002012,superCDMS2014,LUX2014}.

It is also well known that, although the Standard Model (SM) has been
tremendously successful in describing electroweak and strong interaction
phenomena, it must be extended. Physics beyond the SM has both theoretical
and experimental motivations. For instance, the neutrino masses and
mixing, which are required for giving a consistent explanation for
the solar and atmospheric neutrino anomalies, is one of the most firm
evidences to go beyond the SM. Another motivation is providing a satisfactory
explanation to the nature of the DM. This last reason is the focus
of our work. The preferred theoretical framework which provides a
DM candidate is supersymmetry ~\cite{goldberg1983,ellis1984,kane1994,edsjo1997}.
However, many other interesting scenarios have been proposed ~\cite{hooper2007,agashe2004,birkedal2006,burgess2001,barger2009,okada2010,lindner2011}.
In this paper, we focus on the possibility of having a viable scalar
DM candidate in a $U\left(1\right)$ gauge extension of the SM. In
particular, this model, sometimes referred as the flipped $B-L$ model~\cite{montero2009,montero2011}
has a very rich scalar content, which allows us to obtain a complex
scalar DM candidate.

The outline of this paper is the following. In Sec.~\ref{sec:model}
we briefly summarize the model under consideration. In Sec.~\ref{secIII}
we study the vacuum structure and the scalar sector spectrum that
allows us to have a viable complex scalar DM candidate in the model.
In particular, we considered the scalar potential in the context of
$\zeta\equiv\frac{V_{\Phi}}{V_{\phi}}\ll1$, where $V_{\Phi}$ and
$V_{\phi}$ are the vacuum expectation values, VEVs, of the doublets
$\Phi_{1,2}$ and the singlets $\phi_{1,3,X}$ respectively. In Sec.~\ref{secIV}
we present the gauge sector and choose some parameters that simplify
the study of the DM candidates. In Sec.~\ref{secV} we calculate
the thermal relic density of the complex scalar DM candidate and present
a set of parameters that are consistent with the current observations.
In Sec.~\ref{secVI} we summarize the main features of our study.
Finally, in the Appendix, we show the general minimization conditions
used to calculate the scalar mass spectrum.

\section{Brief review of the $B-L$ model}

\label{sec:model} We briefly summarize here the model in Refs.~\cite{montero2009,montero2011}.
It is an extension of the SM based on the gauge symmetry $SU(2)_{L}\otimes U(1)_{Y^{\prime}}\otimes U(1)_{B-L}$
where $B$ and $L$ are the usual baryonic and leptonic numbers, respectively,
and $Y^{\prime}$ is a new charge different from the hypercharge $Y$
of the SM. The values of $Y^{\prime}$ are chosen to obtain the hypercharge
$Y$ through the relation $Y=[Y^{\prime}+(B-L)]$, after the first
spontaneous symmetry breaking. Assuming a generation-independent charge
assignment, the non-existence of mirror fermions and restricting ourselves
to integer quantum numbers for the $Y'$ charge, the anomaly cancellation
constrains the number of right-handed neutrinos, $n_{R}\geq3$~\cite{montero2009}.
Considering $n_{R}=3$, there is an exotic charge assignment for the
$Y'$ charge where $Y'_{n_{R1},n_{R2}}=-4$ and $Y'_{n_{R3}}=5$ besides
the usual one where $Y'_{n_{Ri}}=1$ with $i=1,\,2,\,3$. The model
under consideration has that exotic $Y'$ charge assignment. The respective
fermionic charge assignment of the model is shown in Table\,\ref{table1}.

\begin{table}[h] 
\begin{eqnarray*} 
\begin{array}{|c| c c c c c|} 
\hline \textrm{Fermion} & I_{3} & I & Q & Y^{\prime } & B-L  \\ 
\hline
\hline 
\nu _{eL},\,e_{L} & 1/2,-1/2 & 1/2 & 0,-1 & 0 & -1 \\ 
e_{R} & 0 & 0 & -1 & -1 & -1  \\ 
u_{L},\,d_{L} & 1/2,-1/2 & 1/2 & 2/3,-1/3 & 0 & 1/3  \\
u_{R} & 0 & 0 & 2/3 & 1 & 1/3  \\
d_{R} & 0 & 0 & -1/3 & -1 & 1/3  \\
n_{1R}, n_{2R}  & 0 & 0 & 0 & 4 & -4  \\
n_{3R} & 0 & 0 & 0 & -5 & 5 \\
\hline
\end{array}
\end{eqnarray*}
\caption{Quantum number assignment for the fermionic fields.} 
\label{table1} 
\end{table}

In the scalar sector the model has three $SU(2)_{L}$ doublets, $H,\,\Phi_{1},\,\Phi_{2}$,
and four $SU(2)_{L}$ singlets, $\phi_{1},\,\phi_{2},\,\phi_{3},\,\phi_{X}$.
The scalar charge assignments are shown in Table\,\ref{table2}.
The $H$ doublet is introduced to give mass to the lighter massive
neutral vector boson $Z_{1\mu},$ the charged vector bosons $W_{\mu}^{\pm}$,
and the charged fermions, as in the SM. Besides giving mass to the
extra neutral vector boson $Z_{2\mu}$, which is expected to be heavier
than $Z_{1\mu}$, the other scalars are mainly motivated to generate
mass for both the left and the right handed neutrinos. In order to
be more specific, the other $\Phi_{1}$ and $\Phi_{2}$ doublets are
introduced to give Dirac mass terms at tree level through the renormalizable
Yukawa interactions $\mathcal{D}_{im}\overline{L}_{Li}n_{Rm}\Phi_{1}$
and $\mathcal{D}_{i3}\overline{L}_{Li}n_{R3}\Phi_{2}$ in the Lagrangian.
The $\phi_{1},\,\phi_{2}$ and $\phi_{3}$ singlets are introduced
to generate Majorana mass terms at tree level ($\mathcal{M}_{mn}\overline{(n_{Rm})^{c}}n_{Rn}\phi_{1}$,
$\mathcal{M}_{33}\overline{(n_{R3})^{c}}n_{R3}\phi_{2}$, $\mathcal{M}_{m3}\overline{(n_{Rm})^{c}}n_{R3}\phi_{3}$).
Finally, the $\phi_{X}$ singlet is introduced to avoid dangerous
Majorons when the symmetry is broken down as shown in Ref.~\cite{montero2011}.
These extra scalars allow the model to implement a see-saw mechanism
at $\mathcal{O}\left(\text{TeV}\right)$ energy scale, and the observed
mass-squared differences of the neutrino are obtained without resorting
to fine-tuning the neutrino Yukawa couplings~\cite{montero2011}.
Other studies about the possibility that the model accommodates different
patterns for the neutrino mass matrix using discrete symmetries ($S_{3},\, A_{4}$)
have been done~\cite{dias2012,ana2013}. 

\begin{table}[h] 
\begin{eqnarray*} 
\begin{array}{|c| c c c c c|} 
\hline \textrm{Scalar} & I_{3} & I & Q & Y^{\prime } & B-L \\ 
\hline
\hline 
H^{0,+} & \mp 1/2 & 1/2 & 0,1 & 1 & 0 \\ 
\Phi_{1}^{0,-} & \pm 1/2 & 1/2 & 0,-1 & -4 & 3  \\ 
\Phi_{2}^{0,-} & \pm 1/2 & 1/2 & 0,-1 & 5 & -6  \\ 
\phi_{1} & 0 & 0 & 0 & -8 & 8  \\ 
\phi_{2} & 0 & 0 & 0 & 10 & -10  \\ 
\phi_{3} & 0 & 0 & 0 & 1 & -1  \\ 
\phi_{\textrm{X}} & 0 & 0 & 0 & 3 & -3  \\ 
\hline
\end{array}
\end{eqnarray*}
\caption{Quantum number assignment for the scalar fields.} 
\label{table2} 
\end{table}With the above matter content we can write the most general Yukawa
Lagrangian respecting the gauge invariance as follows
\begin{eqnarray}
-\mathcal{L}_{\text{Y}} & = & Y_{i}^{(l)}\overline{L}_{Li}e_{Ri}H+Y_{ij}^{(d)}\overline{Q}_{Li}d_{Rj}H+Y_{ij}^{(u)}\overline{Q}_{Li}u_{Rj}\widetilde{H}+\mathcal{D}_{im}\overline{L}_{Li}n_{Rm}\Phi_{1}+\mathcal{D}_{i3}\overline{L}_{Li}n_{R3}\Phi_{2}\nonumber \\
 &  & +\mathcal{M}_{mn}\overline{(n_{Rm})^{c}}n_{Rn}\phi_{1}+\mathcal{M}_{33}\overline{(n_{R3})^{c}}n_{R3}\phi_{2}+\mathcal{M}_{m3}\overline{(n_{Rm})^{c}}n_{R3}\phi_{3}+\textrm{H.c.},\label{lyukawa}
\end{eqnarray}
where $i,j=1,2,3$ are lepton/quark family numbers; $m,n=1,2$, and
$\widetilde{H}=i\tau_{2}H^{\ast}$ ($\tau_{2}$ is the Pauli matrix).
Also, we have omitted summation symbols over repeated indices. 

Finally, the most general renormalizable scalar potential obtained
by the addition of all these above mentioned scalar fields is given
by
\begin{eqnarray}
V_{B-L} & = & -\mu_{H}^{2}H^{\dagger}H+\lambda_{H}\left|H^{\dagger}H\right|{}^{2}-\mu_{11}^{2}\Phi_{1}^{\dagger}\Phi_{1}+\lambda_{11}\left\vert \Phi_{1}^{\dagger}\Phi_{1}\right\vert ^{2}-\mu_{22}^{2}\Phi_{2}^{\dagger}\Phi_{2}+\lambda_{22}\left\vert \Phi_{2}^{\dagger}\Phi_{2}\right\vert ^{2}\nonumber \\
 &  & -\mu_{s\alpha}^{2}\left\vert \phi_{\alpha}\right\vert ^{2}+\lambda_{s\alpha}\left\vert \phi_{\alpha}^{\ast}\phi_{\alpha}\right\vert ^{2}+\lambda_{12}\left\vert \Phi_{1}\right\vert ^{2}\left\vert \Phi_{2}\right\vert ^{2}+\lambda_{12}^{\prime}(\Phi_{1}^{\dagger}\Phi_{2})(\Phi_{2}^{\dagger}\Phi_{1})+\Lambda_{H\gamma}\left\vert H\right\vert ^{2}\left\vert \Phi_{\gamma}\right\vert ^{2}\nonumber \\
 &  & +\Lambda_{H\gamma}^{\prime}(H^{\dagger}\Phi_{\gamma})(\Phi_{\gamma}^{\dagger}H)+\Lambda_{Hs\alpha}\left\vert H\right\vert ^{2}\left\vert \phi_{\alpha}\right\vert ^{2}+\Lambda_{\gamma\alpha}^{\prime}\left\vert \Phi_{\gamma}\right\vert ^{2}\left\vert \phi_{\alpha}\right\vert ^{2}+\Delta_{\alpha\beta}(\phi_{\alpha}^{\ast}\phi_{\alpha})(\phi_{\beta}^{\ast}\phi_{\beta})\nonumber \\
 &  & +\left[\beta_{123}\phi_{1}\phi_{2}(\phi_{3}^{\ast})^{2}+\Phi_{1}^{\dagger}\Phi_{2}(\beta_{13}\phi_{1}\phi_{3}^{\ast}+\beta_{23}\phi_{2}^{\ast}\phi_{3})-i\kappa_{H1X}\Phi_{1}^{T}\tau_{2}H\phi_{X}\right.\nonumber \\
 &  & \left.-i\kappa_{H2X}(\Phi_{2}^{T}\tau_{2}H)(\phi_{X}^{\ast})^{2}+\beta_{X}(\phi_{X}^{\ast}\phi_{1})(\phi_{2}\phi_{3})+\beta_{3X}(\phi_{X}^{\ast}\phi_{3}^{3})+\textrm{H.c.}\right],\label{potential}
\end{eqnarray}
where $\gamma=1,2$; $\alpha,\beta=1,2,3,\textrm{X}$; and $\alpha\neq\beta$
in the $\Delta_{\alpha\beta}(\phi_{\alpha}^{\ast}\phi_{\alpha})(\phi_{\beta}^{\ast}\phi_{\beta})$
terms.

\section{The vacuum structure and the scalar sector spectrum\label{secIII} }

In general, DM must be stable to provide a relic abundance in agreement
with the one measured by WMAP and PLANCK, $\Omega_{\textrm{DM}}h^{2}=0.1199\pm0.0027$~\cite{bennettWMAP,planck2013}.
Although the DM stability could result from the extreme smallness
of its couplings to ordinary particles, we restrict ourselves to search
for a discrete, or continuous, symmetry such as $Z_{2}$, or $U(1),$
to protect DM candidates to decay. 

First, we consider the scalar potential in Eq.\,(\ref{potential})
looking for an accidental symmetry that naturally stabilizes the DM
candidate. Doing so, we find that the scalar potential has just the
$SU(2)\otimes U(1)_{Y'}\otimes U(1)_{B-L}$ initial symmetry. However,
none of these gauge groups can generate a stable neutral scalar when
they are spontaneously broken down to $U(1)_{Q}$. Therefore, we impose
a discrete symmetry in the following way: $Z_{2}(\phi_{2})=-\phi_{2}$
and the other scalar fields being even under this $Z_{2}$ symmetry.
As a result, the $\beta_{23}\Phi_{1}^{\dagger}\Phi_{2}\phi_{2}^{\ast}\phi_{3},\,\beta_{123}\phi_{1}\phi_{2}(\phi_{3}^{\ast})^{2}$
and $\beta_{X}(\phi_{X}^{\ast}\phi_{1})(\phi_{2}\phi_{3})$ terms
are prohibited from appearing in the scalar potential, Eq.\,(\ref{potential}).
Actually, when these terms are eliminated from Eq.\,(\ref{potential}),
the true global symmetry in the potential is $SU(2)\otimes U(1)_{Y'}\otimes U(1)_{B-L}\otimes U(1)_{\chi}$,
where the last one is $U(1)_{\chi}:\,\phi_{2}\rightarrow\exp(-i\chi_{\phi_{2}})\phi_{2}$,
where $\chi_{\phi_{2}}$ is the $\phi_{2}$ quantum number under the
$U(1)_{\chi}$ symmetry, and the rest of the fields being invariant.
It is important to say that we have taken into account the simplicity
and some phenomenological criteria to choose the $Z_{2}$ symmetry
above. For example, if we impose $Z_{2}\left(\phi_{1}\right)=-\phi_{1}$
(and the other fields being even), the model has a massless right
handed neutrino, say $N_{R}$, at tree level. That poses a tension
with the experimental data of the invisible $Z_{\mu}$ decay width~\cite{nakamura2010},
since $Z_{\mu}\rightarrow\bar{N_{R}}+N_{R}$ would be allowed to exist~\cite{garcia1989}.
Other simple choices such as $Z_{2}(\phi_{3})=-\phi_{3}$ or $Z_{2}(\phi_{X},\Phi_{1})=-\phi_{X},-\Phi_{1}$
should be avoided due to the appearance of Majorons, $Js$, in the
scalar spectra. As it is well known, the major challenges to models
with Majorons come from the energy loss in stars, through the process
$\gamma+e^{-}\rightarrow e^{-}+J$, and the invisible $Z_{\mu}$ decay
width, through $Z_{\mu}\rightarrow RJ\rightarrow JJJ$, being $R$
a scalar~\cite{mohapatra2004}. 

For the general case of the scalar potential with the $U(1)_{\chi}$
symmetry, we have the minimization conditions given in the Appendix.
In general, those conditions lead to different breaking symmetry patterns
and to a complex vacuum structure because the scalar potential has
several free parameters. In this paper, however, we restrict ourselves
to find a (some) viable scalar DM candidate(s) and to study its (their)
properties in a relevant subset of the parameter space. 

First, we impose the necessary conditions for all real neutral components
of the scalar fields, except $\phi_{2R}$, to obtain nontrivial vacuum
expectation values (VEVs), i.e. $\left\langle H_{R}^{0}\right\rangle =V_{H},\,\left\langle \Phi_{1R}^{0}\right\rangle =V_{\Phi_{1}},\,\left\langle \Phi_{2R}^{0}\right\rangle =V_{\Phi_{2}},\,\left\langle \phi_{1R}\right\rangle =V_{\phi_{1}},\,\left\langle \phi_{2R}\right\rangle =0,\,\left\langle \phi_{3R}\right\rangle =V_{\phi_{3}},\,\left\langle \phi_{XR}\right\rangle =V_{\phi_{X}}$.
For the sake of simplicity, we set $V_{\Phi_{1}}=V_{\Phi_{2}}=V_{\Phi}$
and $V_{\phi_{1}}=V_{\phi_{3}}=V_{\phi_{X}}=V_{\phi}$. Thus, the
$U(1)_{\chi}$ symmetry is not spontaneously broken and the model
possesses two neutral stable scalars which are the real (CP-even)
and the imaginary (CP-odd) parts of the $\phi_{2}$ field with the
same mass given by
\begin{eqnarray}
M_{\textrm{DM}}^{2} & = & \frac{1}{2}\left[\Lambda_{Hs2}V_{\textrm{SM}}^{2}+(\Lambda_{12}^{'}+\Lambda_{22}^{'}-2\Lambda_{Hs2})V_{\Phi}^{2}+(\Delta_{12}+\Delta_{23}+\Delta_{2X})V_{\phi}^{2}-2\mu_{s2}^{2}\right];\label{dmmass}
\end{eqnarray}
where we have defined $V_{\textrm{SM}}^{2}\equiv V_{H}^{2}+V_{\Phi_{1}}^{2}+V_{\Phi_{2}}^{2}=V_{H}^{2}+2V_{\Phi}^{2}=(246)^{2}\textrm{ GeV\ensuremath{^{2}}}$.
From here on, we work with $M_{\textrm{DM}}^{2}$ as an input parameter,
thus we solve Eq.\,(\ref{dmmass}) for $\mu_{s2}^{2}$
\begin{eqnarray}
\mu_{s2}^{2} & = & \frac{1}{2}\left[\Lambda_{Hs2}V_{\textrm{SM}}^{2}+(\Lambda_{12}^{'}+\Lambda_{22}^{'}-2\Lambda_{Hs2})V_{\Phi}^{2}+(\Delta_{12}+\Delta_{23}+\Delta_{2X})V_{\phi}^{2}-2M_{\textrm{DM}}^{2}\right].
\end{eqnarray}
If we allow $\left\langle \phi_{2}\right\rangle \neq0$, the real
part of the $\phi_{2}$ field obtains mass and its imaginary part
is massless and stable. In that case, the DM candidate would be the
Goldstone boson related to the breakdown of the $U(1)_{\chi}$ symmetry.
In general, such massless DM has severe constraints from the big bang
nucleosynthesis~\cite{barger2003,cyburt2005} and the bullet cluster~\cite{randall2008,barger2009}.
Here we do not consider this case. 

Also, we work in the context of $\zeta\equiv\frac{V_{\Phi}}{V_{\phi}}\ll1$.
This assumption allows us to implement a stable and natural see-saw
mechanism for neutrino masses at low energies, as shown in Ref.~\cite{montero2011}.
Once $V_{H}^{2}+2V_{\Phi}^{2}=(246)^{2}\textrm{ GeV\ensuremath{^{2}}}$
and $V_{H}$ is the mainly responsible to give the top quark mass
at tree level, we have $V_{H}^{2}\gg V_{\Phi}^{2}$. Choosing $V_{\phi}\sim1$
TeV and $V_{\Phi}\sim1$ MeV, as in Ref.~\cite{montero2011}, we have
that the $\zeta$ parameter is $\sim10^{-6}$. 

In general, we solve numerically the minimization conditions, and
using standard procedures we construct numerically the mass-squared
matrices for the charged, CP-even and CP-odd scalar fields. We choose
the parameters in the potential such that they satisfy simultaneously
the minimization conditions, the positivity of the squared masses
and the lower boundedness of the scalar potential. In order to satisfy
this last condition, we choose the parameters such that the quartic
terms in the scalar potential are positive for all directions. Although,
all those constraints are checked numerically, let us give an insight
into some constraints coming from the minimization conditions and
the positivity of the squared masses when we do some simplifying assumption
on the parameters. First, we solve the Eqs.\,(\ref{vin1}) and (\ref{vin2})
in the limit $\zeta\rightarrow0$. Doing so, we have
\begin{eqnarray}
\mu_{H} & =\pm & \sqrt{\lambda_{H}V_{\textrm{SM}}^{2}+\frac{1}{2}(\Lambda_{Hs1}+\Lambda_{Hs3}+\Lambda_{HsX})V_{\phi}^{2}}+\mathcal{O}\left(\zeta\right);\label{muh}\\
\kappa_{H1X} & = & \mathcal{O}\left(\zeta\right);\ \ \ \ \ \kappa_{H2X}=\mathcal{O}\left(\zeta\right);\label{kappa1}\\
\mu_{s1} & =\pm & \frac{\sqrt{\Lambda_{Hs1}V_{\textrm{SM}}^{2}+(\Delta_{13}+\Delta_{1X}+2\lambda_{s1})V_{\phi}^{2}}}{\sqrt{2}}+\mathcal{O}\left(\zeta\right);\\
\mu_{s3} & =\pm & \frac{\sqrt{\Lambda_{Hs3}V_{\textrm{SM}}^{2}+(3\beta_{3X}+\Delta_{13}+\Delta_{3X}+2\lambda_{s3})V_{\phi}^{2}}}{\sqrt{2}}+\mathcal{O}\left(\zeta\right);\\
\mu_{sX} & = & \pm\frac{\sqrt{\Lambda_{HsX}V_{\textrm{SM}}^{2}+(\beta_{3X}+\Delta_{1X}+\Delta_{3X}+2\lambda_{sX})V_{\phi}^{2}}}{\sqrt{2}}+\mathcal{O}\left(\zeta\right);
\end{eqnarray}
From Eq.\,(\ref{kappa1}), we see that $\kappa_{H1X}\rightarrow0$
and $\kappa_{H2X}\rightarrow0$ when $\zeta\rightarrow0$ (and keeping
$V_{\phi}$ finite). Thus, in our calculations we choose $\kappa_{H1X}\sim V_{\Phi}$
and $\kappa_{H2X}\sim V_{\Phi}/V_{\phi}$. $ $

To simplify the squared masses and obtain useful analytical expressions,
let us consider $\lambda_{11}=\lambda_{22}=\lambda_{s1}=\lambda_{s3}=\lambda_{sX}$;
$\Lambda_{H1}=\Lambda_{H2}=\Lambda_{Hs1}=\Lambda_{Hs3}=\Lambda_{HsX}=\Lambda_{H1}^{'}=\Lambda_{H2}^{'}$;
$\Lambda{}_{11}^{'}=\Lambda{}_{13}^{'}=\Lambda{}_{1X}^{'}=\Lambda{}_{21}^{'}=\Lambda{}_{23}^{'}=\Lambda{}_{2X}^{'}=\lambda_{12}=\lambda{}_{12}^{'}=\Delta_{13}=\Delta_{1X}=\Delta_{3X}$;
$\Lambda{}_{12}^{'}=\Lambda{}_{22}^{'}=\Delta_{12}=\Delta_{23}=\Delta_{2X}$
and the other parameters without restrictions. The previous constraints
have been inspired by the similitude of the respective potential terms.
We have left free the parameters that involve the DM candidates. Also,
we have assumed that the $H$ scalar field is the Higgs-like field
in this model. Doing these considerations, we have, apart from the
Goldstone bosons that are eaten by the $W^{\pm}$ bosons, two charged
scalars, $C_{1,2}^{\pm}$, with masses given by
\begin{eqnarray}
m_{C_{1}^{\pm},C_{2}^{\pm}}^{2} & = & \frac{1}{4}\left[2\Lambda_{H1}V_{\textrm{SM}}^{2}+\left(1+\sqrt{2}\right)V_{\textrm{SM}}V_{\phi}\right.\nonumber \\
 &  & \left.\mp V_{\phi}\left(\sqrt{\left(3-2\sqrt{2}\right)V_{\textrm{SM}}^{2}+4\beta_{13}^{2}V_{\phi}^{2}}+2\beta_{13}V_{\phi}\right)\right]+\mathcal{O}\left(\zeta\right);\label{charged}
\end{eqnarray}
In the CP-odd scalar sector, we have, besides the two Goldstone bosons
which give mass to the $Z_{1\mu}$ and $Z_{2\mu}$ gauge bosons, the
following mass eigenvalues:
\begin{eqnarray}
m_{I_{3}}^{2} & = & \mathcal{O\left(\zeta\right)};\ \ m_{I_{4}}^{2}=M_{\textrm{DM}}^{2};\ m_{I_{7}}^{2}=-5\beta_{3X}V_{\phi}^{2}+\mathcal{O}\left(\zeta\right);\\
m_{I_{5},I_{6}}^{2} & = & \frac{1}{4}V_{\phi}\left[\left(1+\sqrt{2}\right)V_{\textrm{SM}}-2\beta_{13}V_{\phi}\mp\sqrt{4\beta_{13}^{2}V_{\phi}^{2}+\left(3-2\sqrt{2}\right)V_{\textrm{SM}}^{2}}\right]+\mathcal{O}\left(\zeta\right);
\end{eqnarray}
Finally, in the CP-even scalar sector we have $m_{R_{4}}^{2}=M_{\textrm{DM}}^{2}$,
and
\begin{eqnarray}
m_{R_{5},R_{6}}^{2} & = & \frac{1}{4}V_{\phi}\left[\left(1+\sqrt{2}\right)V_{\textrm{SM}}-2\beta_{13}V_{\phi}\mp\sqrt{4\beta_{13}^{2}V_{\phi}^{2}+\left(3-2\sqrt{2}\right)V_{\textrm{SM}}^{2}}\right]+\mathcal{O}\left(\zeta\right);
\end{eqnarray}
the other mass eigenvalues are not shown for shortness. As shown in
the above expressions in the $\mathcal{O}\left(\zeta\right)$ we have
three degenerate mass eigenstates, i.e. $m_{R_{4}}^{2}=m_{I_{4}}^{2}$,
$m_{R_{5}}^{2}=m_{I_{5}}^{2}$ and $m_{R_{6}}^{2}=m_{I_{6}}^{2}$.
Imposing that all these masses are positive, we find the following
conditions:
\begin{eqnarray}
 & M_{\textrm{DM}}>0\,\land\,\beta_{3X}<0\nonumber \\
 & \land\left[\left(\Lambda_{H2}^{'}>0\land\beta_{13}V_{\phi}+\sqrt{2}V_{\textrm{SM }}<2V_{\textrm{SM }}\right)\lor\right.\nonumber \\
 & \left.\left(V_{\phi}>-2\left(\sqrt{2}-1\right)\Lambda_{H2}^{'}V_{\textrm{SM }}\land\beta_{13}<\frac{V_{\textrm{SM }}(\Lambda_{H2}^{'}V_{\textrm{SM }}+V_{\phi})\left(\Lambda_{H2}^{'}V_{\textrm{SM }}+\sqrt{2}V_{\phi}\right)}{V_{\phi}^{2}\left(2\Lambda_{H2}^{'}V_{\textrm{SM }}+\left(1+\sqrt{2}\right)V_{\phi}\right)}\land\Lambda_{H2}^{'}\leq0\right)\right].\label{conditions}
\end{eqnarray}

Despite the fact that the Eqs.\,(\ref{muh}-\ref{conditions}) are
only valid in the limit $\zeta\rightarrow0,$ these relations will
be useful in our analysis, at least as a starting point.

\section{Gauge Bosons\label{secIV}}

In this model the gauge symmetry breaking proceeds in two stages.
In the first stage, the real components of the $\phi_{1},\,\phi_{3},\,\phi_{X}$
fields obtain VEVs, say $V_{\phi}$, as discussed in the previous
section. Once this happens, the gauge symmetry is broken down to $SU(2)_{L}\otimes U(1)_{Y}$,
where $Y$ is the usual hypercharge of the SM. In the second stage,
the electrically neutral components of the $H,\,\Phi_{1,2}$ obtain
VEVs, $V_{H}$ and $V_{\Phi}$, respectively, thus, breaking down
the symmetry to $U\left(1\right)_{Q}$. 

The mass terms for the three electrically neutral $SU(2)_{L}\otimes U(1)_{Y^{\prime}}\otimes U(1)_{B-L}$
gauge bosons ($W_{\mu}^{3}$, $B_{\mu}^{Y'}$, and $B_{\mu}^{B-L}$)
arise from the kinetic terms for the scalar fields upon replacing
$H,\,\Phi_{1,2},\,\phi_{1,2,3,X}$ by their respective VEVs $\left(\left\langle \phi_{2R}\right\rangle =0\right)$.
In general the mass-square matrix for $W_{\mu}^{3}$, $B_{\mu}^{Y'}$,
and $B_{\mu}^{B-L}$ can be written as follows:
\begin{eqnarray}
{\cal M}_{\textrm{Gauge Bosons }}^{2} & = & \left[\begin{array}{ccc}
g^{2}\left(K+P+2N\right) & -gg_{Y'}\left(K+N\right) & -gg_{B-L}\left(P+N\right)\\
-gg_{Y'}\left(K+N\right) & g_{Y'}^{2}K & g_{Y'}g_{B-L}N\\
-gg_{B-L}\left(P+N\right) & g_{Y'}g_{B-L}N & g_{B-L}^{2}P
\end{array}\right];
\end{eqnarray}
where $g$, $g_{Y'}$ and $g_{B-L}$ are the $SU\left(2\right)_{L},\, U\left(1\right)_{Y'},\, U\left(1\right)_{B-L}$
coupling constants, respectively. $K$, $P$, $N$ are defined by
$K\equiv\frac{1}{4}\sum_{a}V_{a}^{2}Y_{a}^{'2},\ P\equiv\frac{1}{4}\sum_{a}V_{a}^{2}(B-L)_{a}^{2},\ N\equiv\frac{1}{4}\sum_{a}V_{a}^{2}Y'_{a}(B-L)_{a};$
with $Y'_{a}$ and $(B-L)_{a}$ being the quantum numbers given in
the Tables\,\ref{table1} and \ref{table2}. Considering our aforementioned
assumptions we have:
\begin{equation}
K=\frac{1}{4}\left(V_{H}^{2}+41V_{\Phi}^{2}+74V_{\phi}^{2}\right),\, P=\frac{1}{4}\left(45V_{\Phi}^{2}+74V_{\phi}^{2}\right),\, N=-\frac{1}{4}\left(42V_{\Phi}^{2}+74V_{\phi}^{2}\right).
\end{equation}
In order to obtain the relation between the neutral gauge bosons $\left(W_{\mu}^{3},\, B_{\mu}^{Y'},\, B_{\mu}^{B-L}\right)$
and the corresponding mass eigenstates, we diagonalize ${\cal M}_{\textrm{Gauge Bosons }}^{2}$.
Doing so, we have:
\begin{eqnarray}
\gamma_{\mu} & = & \frac{1}{N_{\gamma}}\left[\frac{1}{g}W_{\mu}^{3}+\frac{1}{g_{Y'}}B_{\mu}^{Y'}+\frac{1}{g_{B-L}}B_{\mu}^{B-L}\right];\label{photon}\\
Z_{1\mu} & = & \frac{1}{N_{Z_{1}}}\left[g\left(Pg_{B-L}^{2}-Ng_{Y'}^{2}-M_{Z_{1}}^{2}\right)W_{\mu}^{3}-g_{Y'}\left(\left(P+N\right)g^{2}+Pg_{B-L}^{2}-M_{Z_{1}}^{2}\right)B_{\mu}^{Y'}\right.\nonumber \\
 &  & \left.+g_{B-L}\left(\left(P+N\right)g^{2}+Ng_{Y'}^{2}\right)B_{\mu}^{B-L}\right];\\
Z_{2\mu} & = & \frac{1}{N_{Z_{2}}}\left[g\left(Pg_{B-L}^{2}-Ng_{Y'}^{2}-M_{Z_{2}}^{2}\right)W_{\mu}^{3}-g_{Y'}\left(\left(P+N\right)g^{2}+Pg_{B-L}^{2}-M_{Z_{2}}^{2}\right)B_{\mu}^{Y'}\right]\nonumber \\
 &  & \left.+g_{B-L}\left(\left(P+N\right)g^{2}+Ng_{Y'}^{2}\right)B_{\mu}^{B-L}\right];
\end{eqnarray}
where $N_{\gamma}$, $N_{Z_{1}}$, $N_{Z_{2}}$ are the corresponding
normalization constants. Also, $\gamma_{\mu}$ corresponds to the
photon, and $Z_{1\mu}$ and $Z_{2\mu}$ are the two massive neutral
vector bosons of the model, and their squared masses are given by
$M_{\gamma}^{2}=0$, and 
\begin{eqnarray}
M_{Z_{1\mu},Z_{2\mu}}^{2} & = & \frac{1}{2}R\mp\frac{1}{2}\left[R^{2}-4\left(KP-N^{2}\right)\left(g^{2}\left(g_{Y'}^{2}+g_{B-L}^{2}\right)+g_{Y'}^{2}g_{B-L}^{2}\right)\right]^{1/2},\label{mzeta2}
\end{eqnarray}
with $R\equiv(K+P+2N)g^{2}+Kg_{Y'}^{2}+Pg_{B-L}^{2}$.

For future discussion, it is convenient to define the following basis
\begin{eqnarray}
Z_{\mu} & = & \cos\theta_{W}\, W_{\mu}^{3}-\sin\theta_{W}\,\sin\alpha\, B_{\mu}^{Y'}-\sin\theta_{W}\,\cos\alpha\, B_{\mu}^{B-L};\label{zeta}\\
Z'_{\mu} & = & \cos\alpha\, B_{\mu}^{Y'}-\sin\alpha\, B_{\mu}^{B-L};\label{zetap}
\end{eqnarray}
and the $\gamma_{\mu}$ defined as in Eq.\,(\ref{photon}). The $\alpha$
angle defined as $\tan\alpha\equiv g_{B-L}/g_{Y'}$, can be understood
as the parameter of a particular $SO\left(2\right)$ transformation
on the two gauge bosons, $B_{\mu}^{Y'}$ and $B_{\mu}^{B-L}$, that
rotates the $U\left(1\right)_{Y'}\otimes U\left(1\right)_{B-L}$ gauge
group into the $U\left(1\right)_{Y}\otimes U\left(1\right)_{Z}$ gauge
group. In the last expression $U\left(1\right)_{Y}$ is the usual
hypercharge gauge group. Also, we have that $g^{2}\sin^{2}\theta_{W}=e^{2}=\left(1/g^{2}+1/g_{Y'}^{2}+1/g_{B-L}^{2}\right)^{-1}$.
The $U\left(1\right)_{Z}$ can be understood as the gauge group with
the coupling $g_{Z}^{2}=g_{Y'}^{2}+g_{B-L}^{2}$. Using Eqs.\,(\ref{zeta})
and (\ref{zetap}), we can write the two massive gauge bosons $Z_{1\mu}$
and $Z_{2\mu}$ in terms of $Z_{\mu}$ and $Z'_{\mu}$ as follows:
\begin{eqnarray}
Z_{1\mu} & = & \cos\beta\, Z_{\mu}+\sin\beta\, Z'_{\mu};\label{zeta1m}\\
Z_{2\mu} & = & -\sin\beta\, Z_{\mu}+\cos\beta\, Z'_{\mu};\label{zeta2m}
\end{eqnarray}
where 
\begin{eqnarray}
\tan\beta & = & \frac{\sqrt{g^{2}\left(g_{Y'}^{2}+g_{B-L}^{2}\right)+g_{Y'}^{2}g_{B-L}^{2}}\left(g_{B-L}^{2}P-g_{Y'}^{2}N-M_{Z_{2}}^{2}\right)}{g^{2}\left(g_{Y'}^{2}+g_{B-L}^{2}\right)(P+N)+g_{Y'}^{2}\left(g_{B-L}^{2}(P+N)-M_{Z_{2}}^{2}\right)}.
\end{eqnarray}
From Eqs.\,(\ref{mzeta2}), (\ref{zeta1m}) and (\ref{zeta2m}),
we can see that $\tan\beta=0$ when $V_{\phi}\rightarrow\infty$ or
$V_{H}^{2}=\left(g_{Y'}^{2}+3g_{B-L}^{2}\right)V_{\Phi}^{2}/g_{Y'}^{2}$.
However, this last solution is not allowed since in our case we have
$V_{H}\gg V_{\Phi}$ and ${\cal O}\left(g_{Y'}\right)\sim{\cal O}\left(g_{B-L}\right)$.

In this work, we use the following gauge couplings, $g\simeq0.65,\, g_{Y'}=g_{B-L}\simeq0.505$,
such that $\tan\beta\simeq4\times10^{-4}$. Doing so, we have $Z_{1\mu}\simeq Z_{\mu}$
and $Z_{2\mu}\simeq Z'_{\mu}$. In general, the $\beta$ angle must
be quite small, $\beta\lesssim10^{-3}$, to be in agreement with precision
electroweak studies~\cite{erler2009,aguila2010,diener2012} since
a new neutral boson $Z_{2\mu}$ which mixes with the SM $Z_{\mu}$
distorts its properties, such as couplings to fermions and mass relative
to electroweak inputs. Using those parameters for the gauge couplings
and the VEVs discussed in the previous section, we obtain $M_{Z'}\simeq3.1$
TeV besides the already known masses for the SM gauge bosons. In general,
a new neutral vector boson must have a mass in the order of few TeV,
or be very weakly coupled to the known matter to maintain consistency
with the present phenomenology~\cite{appelquist2003,carena2004,erler2009,aguila2010,diener2012,han2013}.
Doing a phenomenological study of the bounds on the parameter space
imposed by data coming from LEP II, Tevatron and LHC in the present
model is out of the scope of this work. However, we see that the $M_{Z'}$
value above is consistent with the relation $M_{Z'}/g_{B-L}\simeq6.13\gtrsim6$
TeV ~\cite{appelquist2003,carena2004}.

Finally, the charged gauge bosons $W_{\mu}^{\pm}$ are not affected
by the presence of one additional neutral gauge boson $Z_{2\mu}$.
These have the same form as in the SM, $W_{\mu}^{\pm}=\frac{1}{\sqrt{2}}\left(W_{\mu}^{1}\mp iW_{\mu}^{2}\right)$,
with masses given by $M_{W^{\pm}}^{2}=\frac{1}{4}g^{2}V_{\textrm{SM}}^{2}=\frac{1}{4}g^{2}\left(V_{H}^{2}+2V_{\Phi}^{2}\right)$.

\section{Dark Matter\label{secV}}

\subsection{Thermal Relic Density }

In order to calculate the present-day DM mass density, $\Omega_{\textrm{DM}}h^{2}$,
arising from $R_{\textrm{DM}}$ and $I_{\textrm{DM}}$ scalars freezing
out from thermal equilibrium, we follow the standard procedure in
Refs.~\cite{gondolo1991,griest1991}. Thus, we should find the solution
to the Boltzmann equations for the $Y_{R_{\textrm{\textrm{DM}}}}$
and $Y_{I_{\textrm{DM}}}$, which are defined as the ratio of the
number of particles ($n_{R_{\textrm{DM }}}$ and $n_{I_{\textrm{DM}}}$)
to the entropy, $Y_{i}\equiv n_{i}/s$ ($i=R_{\textrm{DM}},\, I_{\textrm{DM}}$),
with $s$ being the total entropy density of the Universe. Usually,
$s$ is written in terms of the effective degrees of freedom $h_{\textrm{eff}}\left(T\right)$
as follows: $s=\frac{2\pi^{2}}{45}h_{\textrm{eff}}\left(T\right)T^{3};$
where $T$ is the photon temperature and $h_{\textrm{eff}}$ is calculated
as in the Ref.~\cite{gondolo1991}. Actually, in our case, due to
the $U(1)_{\chi}$ symmetry introduced in Sec.~\ref{secIII}, $M_{I_{\textrm{DM}}}=M_{R_{\textrm{DM}}}=M_{\textrm{DM}}$,
$Y_{R_{\textrm{DM}}}=Y_{I_{\textrm{DM}}}\equiv Y$, and, $\Omega_{\textrm{DM}}h^{2}=\Omega_{R_{\textrm{DM}}}h^{2}+\Omega_{I_{\textrm{DM}}}h^{2}=2\Omega_{R_{\textrm{DM}}}h^{2}=2\Omega_{I_{\textrm{DM}}}h^{2}$.
Therefore, the Boltzmann equation that we have to solve is 
\begin{eqnarray}
\frac{dY}{dx} & = & -\left(\frac{45}{\pi}G\right)^{-1/2}\frac{g_{*}^{1/2}M_{\textrm{DM}}}{x^{2}}\left\langle \sigma v_{\textrm{Moller}}\right\rangle _{\textrm{ann }}\left[Y^{2}-Y_{\textrm{eq}}^{2}\right];\label{BoltzmannEq}
\end{eqnarray}
where $x=M_{\textrm{DM}}/T$, $G$ is the gravitational constant,
and $Y_{\textrm{eq}}=n_{\textrm{eq}}/s$. $n_{\textrm{eq}}$ is the
thermal equilibrium number density and when $M_{\textrm{DM}}/T\gg1$,
it is $n_{\textrm{eq}}=g_{i}\left(\frac{M_{\textrm{DM}}T}{2\pi}\right)^{3/2}\exp\left[-\frac{M_{\textrm{DM}}}{T}\right];$
where $g_{i}=1$ is the internal degree of freedom for the scalar
dark matter. The $g_{*}$ parameter in the Eq.\,(\ref{BoltzmannEq})
is calculated as in the Ref.~\cite{gondolo1991}. 

Also, we have that the thermal-average of the annihilation cross section
times the Moller velocity, $\left\langle \sigma v_{\textrm{Moller}}\right\rangle _{\textrm{ann}}$,
has the following form
\begin{eqnarray}
\left\langle \sigma v_{\textrm{Moller}}\right\rangle _{\textrm{ann}} & = & \frac{1}{8M_{\textrm{DM}}^{4}TK_{2}^{2}\left(M_{\textrm{DM}}/T\right)}\int_{4M_{\textrm{DM}}^{2}}^{\infty}\sigma_{\textrm{ann}}\left(s-4M_{\textrm{DM}}^{2}\right)\sqrt{s}K_{1}\left(\sqrt{s}/T\right)ds,
\end{eqnarray}
where $K_{i}$ are the modified Bessel functions of order $i$. The
variable $s$, in the integral above, is the Mandelstam variable.
Finally, once the $Y$ is numerically calculated for the present time,
$Y_{0}$, we can obtain $\Omega_{\textrm{DM}}h^{2}=2.82\times10^{8}\times\left(2\times Y_{0}\right)\times\frac{M_{\textrm{DM}}}{\textrm{GeV}}$
.

\begin{figure}[tbh]
\includegraphics[scale=0.75]{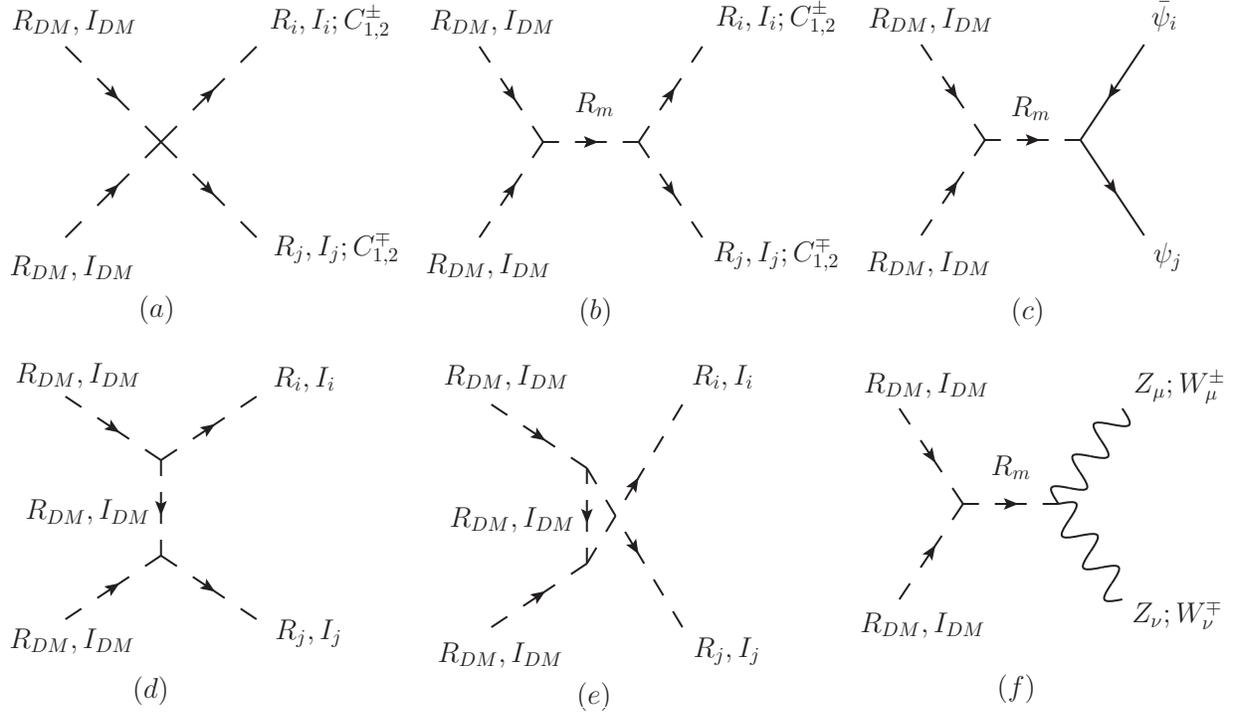}

\caption{Main annihilation processes that contribute to $ $$\left\langle \sigma v_{\textrm{Moller}}\right\rangle _{\textrm{ann}}$.\label{annihilation processes}}
\end{figure}

In order to calculate $\sigma_{\textrm{ann}}$, we have taken into
account all dominant annihilation processes which are shown in Fig.
(\ref{annihilation processes}). In our case, the dominant annihilation
contributions come from the scalar exchange. This is due to the fact
that our DM candidates, $R_{DM}$ and $I_{DM}$, couple neither to
$Z_{\mu}$ nor to $W_{\mu}^{\pm}$ gauge bosons at tree level, since
they are SM singlets. Also, we have found that contributions coming
from $Z'_{\mu}$ exchange are negligible for the parameter set considered
here. 

\begin{figure}[tbh]
\includegraphics{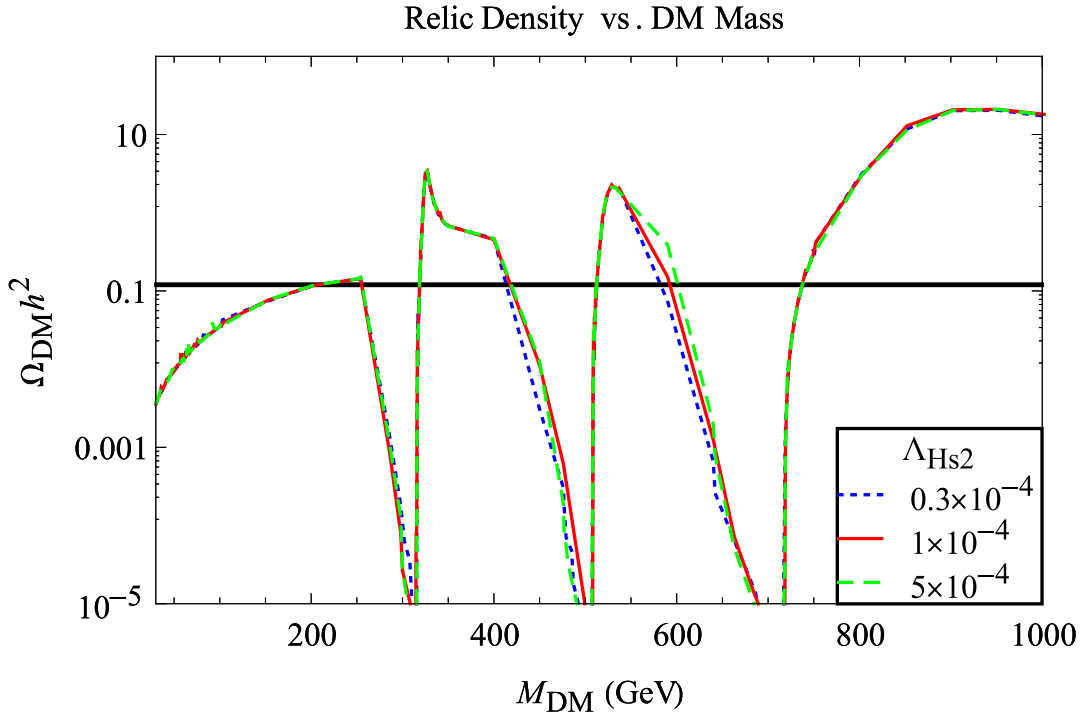}\caption{The total thermal relic density of the $I_{\textrm{DM}}$ and $R_{\textrm{DM}}$
as a function of $M_{\textrm{DM}}$. We have used three different
parameters for $\Lambda_{Hs2}=0.3\times10^{-4},\,1\times10^{-4},\,5\times10^{-4}$.
\label{Fig2}}
\end{figure}

Taking into account all previously mentioned, we have solved numerically
Eq.\,(\ref{BoltzmannEq}) for a representative set of parameters.
Although the scalar potential in this model has many free parameters,
we have found that the most relevant parameters in determining the
correct DM relic density and in satisfying the currently direct experimental
limits are $\Lambda_{Hs2},\,\Delta_{\alpha2}\,(\textrm{with }\alpha=1,3,X)$
and $\Lambda'_{\gamma2}\,(\textrm{with }\gamma=1,2)$. The $\Lambda_{Hs2}$
coupling strongly controls the direct detection signal, since in our
case both the Higgs-like scalar is almost totally the neutral CP-even
component of the $H$ field and as discussed below, the direct detection
is mainly mediated by the $t-$channel Higgs exchange. In order to
obtain the correct direct detection limits without resorting to resonances,
we found that $\Lambda_{Hs2}\sim10^{-4}$. The $\Delta_{\alpha2}$
and $\Lambda'_{\gamma2}$ parameters are also crucial in obtaining
the correct $\Omega_{\textrm{DM}}h^{2}$ because they mostly control
the $DM-DM-R_{i}\left(I_{i}\right)-R_{i}\left(I_{i}\right)$ and $DM-DM-R_{i}$
couplings and, therefore, the $\sigma_{\textrm{ann}}$. The latter
is not allowed to vary in a wide range since, roughly, $\Omega_{\textrm{DM}}h^{2}\sim1/\left\langle \sigma v_{\textrm{Moller}}\right\rangle _{\textrm{ann}}$
and we aim to obtain values close to $\Omega_{\textrm{DM}}h^{2}\sim0.11$.
In other words, the larger $\Delta_{\alpha2}$ and $\Lambda'_{\gamma2}$
parameters are, the smaller the $\Omega_{\textrm{DM}}h^{2}$ is. In
\,(\ref{Fig2}), we have used $\Lambda'_{\gamma2}\simeq10^{-2}$
and $\Delta_{\alpha2}\simeq9\times10^{-2}$. It is also important
here to mention that the dominant process is the $DM+DM\rightarrow I_{3}+I_{3}$
annihilation, where $I_{3}$ refers to the lightest CP-odd scalars,
as in Sec\,(\ref{secIII}). Although the other parameters in the
scalar potential are not as critical in determining the $\Omega_{\textrm{DM}}h^{2}$,
they give the other quantitative characteristics appearing in Fig.\,(\ref{Fig2}).
In order to be more specific, we have choose the other parameters
such that the mass scalar spectrum is given by: $1437.6,\,1016.9,\,631.7,\,544.9,\,379.6,\,125$
GeV and $707.1,\,544.9,\,379.6,\,2.3\times10^{-6}$ GeV for the CP-even
and CP-odd scalars, respectively. The CP-even scalars with masses
$1437.6,\,1016.9,\,631.7$ GeV have components only in the singlets
$\phi_{1,3,X}$ and the CP-even scalars with $544.9,\,379.6$ GeV
have components only in the scalar doublets $\Phi_{1,2}$. The CP-even
scalar with $125$ GeV has component in the $H$ doublet and it is
the Higgs-like scalar in our model. In Fig.\,(\ref{Fig2}), we can
also observe three resonances in $\simeq315.8,\,508.5,\,718.8$ GeV
corresponding to the $s-$channel exchange of CP-even scalars with
components in the singlets. Let us also mention that the processes
via the $s-$channel due to the exchange of the CP-even scalars with
masses $125,\,379.6,\,544.9$ GeV are highly suppressed because of
the smallness of their couplings. Thus, their resonances do not appear
in Fig.\,(\ref{Fig2}).

\subsection{Direct Detection}

Despite weakly coupled to baryons, WIMPs can scatter elastically with
atomic nuclei, providing the opportunity for direct detection. Currently,
there are several experiments which aim to directly observe WIMP dark
matter~\cite{XENON1002012,superCDMS2014,LUX2014}. The signal in these
experiments is the kinetic energy transferred to a nucleus after it
scatters off a DM particle. The energies involved are less or of the
order of $10$ keV. At these energies the WIMP sees the entire nucleus
as a single unit, with a net mass, charge and spin. In general, the
WIMP-nucleus interactions can be classified as either spin-independent
or spin-dependent. In our case, these interactions are spin-independent
because the two DM candidates are scalars. The relevant WIMP-nucleus
scattering process for direct detection in the case considered here
takes place mainly through the $t-$channel elastic scattering due
to Higgs exchange:$(I_{\textrm{DM}},R_{\textrm{DM}})+N\rightarrow(I_{\textrm{DM}},R_{\textrm{DM}})+N$
($N$ refers to the atomic nucleus). The spin-independent cross section
is given by
\begin{eqnarray}
\sigma_{\chi N}^{\textrm{SI}} & = & \frac{4}{\pi}\frac{M_{\textrm{DM}}^{2}m_{N}^{2}}{\left(M_{\textrm{DM}}+m_{N}\right)^{2}}\left[Zf_{p}+\left(A-Z\right)f_{n}\right]^{2};
\end{eqnarray}
 where the effective couplings to protons and neutrons, $f_{p,n}$,
are
\begin{eqnarray}
f_{p,n} & = & \sum_{q=u,d,s}\frac{G_{\textrm{eff},q}}{\sqrt{2}}f_{Tq}^{\left(p,n\right)}\frac{m_{p,n}}{m_{q}}+\frac{2}{27}f_{TG}^{\left(p,n\right)}\sum_{q=c,b,t}\frac{G_{\textrm{eff},q}}{\sqrt{2}}\frac{m_{p,n}}{m_{q}}.
\end{eqnarray}
By using $f_{Tq}^{\left(p,n\right)}$ and $f_{TG}^{\left(p,n\right)}$
given in Ref.~\cite{ellis2000} and the fact that, in our case, $G_{\textrm{eff},q}=G_{0}\times m_{q}\equiv\frac{C_{DM^{2}H}}{V_{H}M_{\textrm{Higgs}}^{2}}\times m_{q}$
(with $C_{DM^{2}H}$ being the coupling $DM-DM-\textrm{Higgs}$ which
depends on the parameters of the model), we arrive at the cross section
per nucleon of
\begin{equation}
\sigma_{\chi,p}^{\textrm{SI}}\thickapprox2.7\times10^{7}\times\frac{M_{\textrm{DM}}^{2}m_{N}^{2}}{\left(M_{\textrm{DM}}+m_{N}\right)^{2}}\times G_{0}^{2}\textrm{ in pbarn}.
\end{equation}

\begin{figure}[tbh]
\includegraphics{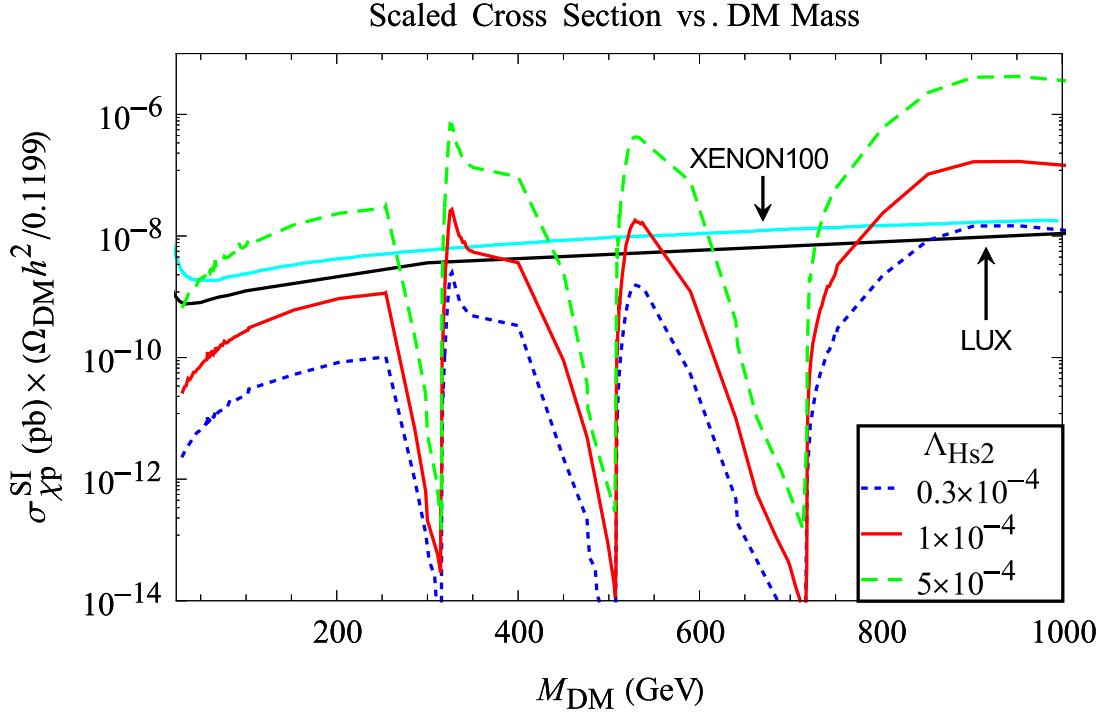}\caption{The spin-independent elastic scattering cross section, $\sigma_{\chi,p}^{\textrm{SI}}$,
off a proton $p$ as a function of $M_{\textrm{DM}}$ for the same
parameters as in Fig.\ref{Fig2}, appropriately scaled to relic density.
We also show the XENON100 and LUX exclusion limits~\cite{XENON1002012,LUX2014}.\label{Fig3}}
\end{figure}

Recently, the Large Underground Xenon (LUX) experiment~\cite{LUX2014}
has reported its first results, setting limits on spin-independent
WIMP-nucleon elastic scattering with a minimum upper limit on the
cross section of $7.6\times10^{-10}$ pbarn at a WIMP mass of $33$
GeV$/c^{2}$. We have found that choosing $\Lambda_{Hs2}\sim10^{-4}$
we obtain the LUX bound without resorting to resonances. It is clear
that values of $\Lambda_{Hs2}$ larger can be considered. However,
we have chosen this conservative value for $\Lambda_{Hs2}$. Our results
are shown in Fig.\,(\ref{Fig3}). The parameters are the same as
in Fig.\,(\ref{Fig2}). 

From Figs.\,(\ref{Fig2}) and (\ref{Fig3}), we see that for a DM
candidate with mass around $200$ GeV and $\Lambda_{Hs2}=0.3\times10^{-4},\,1\times10^{-4},$
the two conditions, $\Omega_{\textrm{DM}}h^{2}$ and the direct detection,
are satisfied outside the resonance regions. We also have verified
that this is a general characteristic of this model. Due to the existence
of the light $I_{3}$ scalar the annihilation process $DM+DM\rightarrow I_{3}+I_{3}$
Fig. (\ref{annihilation processes},a) is the dominant one so that
we do not have to appeal to resonances to get compatibility with experiments.
Other $M_{\textrm{DM }}$values which satisfy the experimental bounds
are shown in Figs\,(\ref{Fig2}) and (\ref{Fig3}). Specifically,
$M_{\textrm{DM}}\thickapprox319,\,410,\,511,\,590,\,737$ GeV are
also possible solutions. However, these are within regions with resonances.

Let us now do some important remarks about the impact of the existence
of $I_{3}$ in this model. First of all, we have a tree level contribution
to the Higgs invisible decay, $\Gamma_{h}^{\textrm{Inv}}$, due to
the coupling of the Higgs field with the light pseudo-scalar field,
$c_{hI_{3}I_{3}}$, which comes from the Lagrangian terms of the form
$\vert H\vert^{2}\vert\phi_{1,2,X}\vert^{2}$, and gives $\Gamma_{hI_{3}I_{3}}^{\textrm{Inv}}=c_{hI_{3}I_{3}}^{2}/32\pi m_{\textrm{h}}$
for $m_{I_{3}}\ll m_{\textrm{h}}$. Actually, when $2M_{\textrm{DM}}<m_{\textrm{h}}$
the $h\rightarrow I_{\textrm{DM}}\, I_{\textrm{DM}}$ and $h\rightarrow R_{\textrm{DM}}\, R_{\textrm{DM}}$
decays are also allowed, thus, further contributing to $\Gamma_{h}^{\textrm{Inv}}$
according to $\Gamma_{h\textrm{DMDM}}^{\textrm{Inv }}=\Gamma_{hI_{\textrm{DM}}I_{\textrm{DM}}}^{\textrm{Inv }}+\Gamma_{hR_{\textrm{DM}}R_{\textrm{DM}}}^{\textrm{Inv }}=2\times c_{h\textrm{DMDM}}^{2}/(32\pi m_{\textrm{h}})\times\sqrt{1-4M_{DM}^{2}/m_{\textrm{h}}^{2}}$
with $c_{h\textrm{DMDM}}\approx\Lambda_{Hs2}V_{H}$. The current limit
on the branching ratio into invisible particles of the Higgs, $BR_{h}^{\textrm{Inv}}$,
is around $10\%-15\%$~\cite{belanger2013,ellis2013}. A stronger
bound of $BR_{h}^{\textrm{Inv}}<5\%$ at $14$ TeV LHC has been claimed~\cite{peskin2013}.
From the set of parameters used to obtain Fig.\,(\ref{Fig2}) and
Fig.\,(\ref{Fig3}) we have that $BR_{h}^{\textrm{Inv}}=\left(\Gamma_{hI_{3}I_{3}}^{\textrm{Inv}}+\Gamma_{h\textrm{DMDM}}^{\textrm{Inv }}\right)/\left(\Gamma_{h}^{\textrm{Vis}}+\Gamma_{hI_{3}I_{3}}^{\textrm{Inv}}+\Gamma_{h\textrm{DMDM}}^{\textrm{Inv }}\right)\simeq3.78\%$
for $M_{\textrm{DM}}=50$ GeV. For different $M_{\textrm{DM}}$ values
we have found $BR_{h}^{\textrm{Inv}}<5\%$. Also, we have used $\Gamma_{h}^{\textrm{Vis}}=4.07$
MeV for $m_{\textrm{H}}=125$ GeV. The model is also safe regarding
the severe existing constraints on the invisible decay width of $Z_{\mu}$
boson since there is no a process like $Z_{\mu}\to RI_{3}\to I_{3}I_{3}I_{3}$~\cite{garcia1989}
due to the fact that $I_{3}$ has only components in the SM singlets.
(It would be kinetically forbidden anyway once all real scalar fields
of the model are heavier than the $Z_{\mu}$ boson.) For the same
reason, there is no issue with the energy loss in stars astrophysical
constraint since there is no tree level coupling inducing the $\gamma+e^{-}\to e^{-}+I_{3}$~\cite{mohapatra2004}.
Finally, some last comments are necessaries. In general, the $I_{3}$
could also contribute to the $\Omega_{\textrm{DM}}h^{2}$ because
it is massive. However, the $I_{3}$ pseudo-scalar is not stable.
It decays mainly in active neutrinos, $\nu$'s, with $\Gamma_{I_{3}\rightarrow\nu\nu}\approx\frac{m_{I_{3}}}{16\pi}\frac{\sum_{i}m_{\nu i}^{2}}{V_{\phi}^{2}}$~\cite{lattanzi2007}.
For the parameter set used here, we have $\tau_{I_{3}}\simeq1/\Gamma_{I_{3}\rightarrow\nu\nu}\approx10^{9}$\,s,
where we have used $\sum_{i}m_{\nu i}^{2}\lesssim0.01$\,eV$^{2}$.
With $\tau_{I_{3}}$ given here and $t_{U}\simeq4.3\times10^{17}$s
(age of the Universe), the $\Omega_{I_{3}}h^{2}\simeq\frac{m_{I_{3}}}{1.25\textrm{\,\ keV}}\exp\left(-t_{\textrm{U}}/\tau_{I_{3}}\right)\simeq0$.
In the last expression for $\Omega_{I_{3}}h^{2}$ we have considered
that the $T_{DI_{3}}>175$ GeV (where $T_{DI_{3}}$ is the decoupling
temperature of the $I_{3}$). There is also a constraint comes from
the observed large scale structure of the Universe~\cite{akhmedov1993,steigman1985}.
Roughly speaking, this last condition impose $r_{I_{3}}\frac{m_{I_{3}}}{1\textrm{ keV}}\left(\frac{\tau_{I_{3}}}{1\textrm{ s}}\right)^{1/2}\lesssim4\times10^{3}$~\cite{akhmedov1993}.
In last expression $r_{I_{3}}=g_{\textrm{eff}}(T_{0})/g_{\textrm{eff}}(T_{DI_{3}})\approx1/25$,
being $g_{\textrm{eff}}$ the effective number of the relativistic
degrees of freedom. With our parameter set this condition is satisfied.$ $

\section{Conclusions\label{secVI}}

We have discussed in this work a scenario where a complex DM candidate
is possible. In particularly, the model studied here is a gauge extension
of the SM based on a $SU(2)_{L}\otimes U(1)_{Y^{\prime}}\otimes U(1)_{B-L}$
symmetry group. This model contents three right handed neutrinos and
some extra scalars, doublets and singlets, with different quantum
numbers. In principle, those scalars are introduced to generate Majorana
and Dirac mass terms at the tree level and to allow the implementation
of a see-saw mechanism at the TeV scale as shown in Ref.~\cite{montero2011}.
The non-standard doublets and singlets introduce two new energy scales,
besides the electoweak one given $V_{H}=246$ GeV: $V_{\Phi}$ (the
VEVs of the extra doublet neutral scalars) and $V_{\phi}$ (the VEVs
of the extra singlet neutral scalars). If $\zeta\equiv V_{\Phi}/V_{\phi}\ll1$
the see-saw mechanism becomes natural~\cite{montero2011}. In this
context, we have studied the scalar spectrum and imposed a $Z_{2}$
symmetry on the $\phi_{2}$ singlet scalar (which accidentally became
a $U(1)_{\chi}\,\textrm{symmetry}:\,\phi_{2}\rightarrow\exp(-i\chi_{\phi_{2}})\phi_{2}$)
in order to allow a complex DM candidate. Before studying the constraints
coming from the thermal relic density ($\Omega_{\textrm{DM}}h^{2}$)
and direct detection experiments on this DM candidate, we have done
a brief analysis of the gauge sector concerning the $Z_{\mu},Z_{\mu}^{\prime}$
mixing angle ($\tan\beta\simeq4\times10^{-4}$) which satisfies the
$\beta\lesssim10^{-3}$ electroweak precision constraint, and we have
verified that the $Z_{\mu}^{\prime}$ mass emerging from the model
is consistent with the relation $M_{Z'}/g_{B-L}\simeq6.13\gtrsim6$
TeV. Then, we have chosen some parameters that simultaneously allow
us to have a compatible $\Omega_{\textrm{DM}}h^{2}$ and satisfy the
direct detection experiments. Although the scalar potential has many
parameters, we have found that the $\Lambda_{Hs2},\,\Delta_{\alpha2}\,(\textrm{with }\alpha=1,3,X)$
and $\Lambda'_{\gamma2}\,(\textrm{with }\gamma=1,2)$ parameters mostly
control these two constraints. The $\Lambda_{Hs2}$ parameter is fundamental
in satisfying the limits coming from direct detection, since in our
case it takes place through the $t-$channel elastic scattering due
to the Higgs exchange. Choosing $\Lambda_{Hs2}\sim10^{-4}$ roughly
satisfies the bounds from the LUX experiment and allows a $\Omega_{\textrm{DM}}h^{2}$
in agreement with the WMAP and PLANCK experiments. The $\Delta_{\alpha2}$
and $\Lambda'_{\gamma2}$ parameters control $\sigma_{\textrm{ann}}$
mostly and, therefore $\Omega_{\textrm{DM}}h^{2}$. As an example,
we have shown $\Omega_{\textrm{DM}}h^{2}$ and $\sigma_{\chi,p}^{\textrm{SI}}$,
for $\Lambda'_{\gamma2}\simeq10^{-2}$ and $\Delta_{\alpha2}\simeq9\times10^{-2}$,
in Figs.\,(\ref{Fig2}) and (\ref{Fig3}). It is interesting to note
that this model, for the same set of parameters fixed, except $M_{\textrm{DM}}$'s,
has several $M_{\textrm{DM}}$ values satisfying the experimental
bounds. In other words, we have found solutions in the region outside
and inside of the resonances for the same parameters, varying $M_{\textrm{DM}}$
only. As previously mentioned, the presence of a light scalar, $I_{3}$,
in this model makes the process $DM+DM\rightarrow I_{3}+I_{3}$ to
be dominant for $\Omega_{\textrm{DM}}h^{2}$. However, $I_{3}$ may
bring some potential problems, so that we have discussed some constraints
imposed on $I_{3}$ coming from the Higgs and the $Z_{\mu}$ invisible
decay widths, the energy loss in stars and the observed large scale
structure of the Universe. We have found that in our context all of
these constraints are satisfied. Finally, we would like to point out
the recent work studying the possibility of having a Majoron DM candidate~\cite{farinaldo2014}.
\begin{acknowledgments}
B.L.S.V. and E.R.S. would like to thank CAPES for financial support
and B.L.S.V. the Argonne National Laboratory for kind hospitality.
We are grateful to E.C.F.S. Fortes, R. Rosenfeld and V. Pleitez for
valuable discussions.
\end{acknowledgments}

\section*{APPENDIX: THE MINIMIZATION CONDITIONS\label{Appendix} }

The general minimization conditions coming from $\partial V_{1}/\partial R_{i}=0$,
where $V_{1}$ is the scalar potential with $U(1)_{\chi}$ symmetry
and $R_{i}=\{H_{R}^{0},\,\Phi_{1R}^{0},\,\Phi_{2R}^{0},\,\phi_{1R},\,\phi_{2R},\,\phi_{3R}\,\phi_{XR}\}$
are the neutral real components of the scalar fields, can be written
as:

\begin{eqnarray}
0 & = & V_{H}\left(2\lambda_{H}V_{H}^{2}+\Lambda_{\text{H1}}V_{\Phi_{1}}^{2}+\Lambda_{\text{H2}}V_{\Phi_{2}}^{2}+\Lambda_{\text{Hs1}}V_{\phi_{1}}^{2}+\Lambda_{\text{Hs2}}V_{\phi_{2}}^{2}+\Lambda_{\text{Hs3}}V_{\phi_{3}}^{2}+\Lambda_{\text{HsX}}V_{\phi_{X}}^{2}-2\mu_{H}^{2}\right)\nonumber \\
 &  & -\sqrt{2}\kappa_{\text{H1X}}V_{\Phi_{1}}V_{\phi_{X}}-\kappa_{\text{H2X}}V_{\Phi_{2}}V_{\phi_{X}}^{2};\label{vin1}\\
0 & = & V_{\Phi_{1}}\left(\Lambda_{\text{H1}}V_{H}^{2}+2\lambda_{11}V_{\Phi_{1}}^{2}+(\lambda'_{12}+\lambda_{12})V_{\Phi_{2}}^{2}+\Lambda'_{11}V_{\phi_{1}}^{2}+\Lambda'_{12}V_{\phi_{2}}^{2}+\Lambda'_{13}V_{\phi_{3}}^{2}+\Lambda'_{1X}V_{\phi_{X}}^{2}-2\mu_{11}^{2}\right)\nonumber \\
 &  & -\sqrt{2}\kappa_{\text{H1X}}V_{H}V_{\phi_{X}}+\beta_{13}V_{\Phi_{2}}V_{\phi_{1}}V_{\phi_{3}};\\
0 & = & V_{\Phi_{2}}\left(\Lambda_{\text{H2}}V_{H}^{2}+(\lambda_{12}+\lambda'_{12})V_{\Phi_{1}}^{2}+2\lambda_{22}V_{\Phi_{2}}^{2}+\Lambda'_{21}V_{\phi_{1}}^{2}+\Lambda'_{22}V_{\phi_{2}}^{2}+\Lambda'_{23}V_{\phi_{3}}^{2}+\Lambda'_{2X}V_{\phi_{X}}^{2}-2\mu_{22}^{2}\right)\nonumber \\
 &  & -\kappa_{\text{H2X}}V_{H}V_{\phi_{X}}{}^{2}+\beta_{13}V_{\Phi_{1}}V_{\phi_{1}}V_{\phi_{3}};\\
0 & = & V_{\phi_{1}}\left(\Lambda_{\text{Hs1}}V_{H}^{2}+\Lambda'_{11}V_{\Phi_{1}}^{2}+\Lambda'_{21}V_{\Phi_{2}}^{2}+2\lambda_{\text{s1}}V_{\phi_{1}}^{2}+\Delta_{12}V_{\phi_{2}}^{2}+\Delta_{13}V_{\phi_{3}}^{2}+\Delta_{1X}V_{\phi_{X}}^{2}-2\mu_{\text{s1}}^{2}\right)\nonumber \\
 &  & +\beta_{13}V_{\Phi_{1}}V_{\Phi_{2}}V_{\phi_{3}};\\
0 & = & V_{\phi_{2}}\left(\Lambda_{\text{Hs2}}V_{H}^{2}+\Lambda'_{12}V_{\Phi_{1}}^{2}+\Lambda'_{22}V_{\Phi_{2}}^{2}+\Delta_{12}V_{\phi_{1}}^{2}+2\lambda_{\text{s2}}V_{\phi_{2}}^{2}+\Delta_{23}V_{\phi_{3}}^{2}+\Delta_{2X}V_{\phi_{X}}^{2}-2\mu_{\text{s2}}^{2}\right);\\
0 & = & V_{\phi_{3}}\left(\Lambda_{\text{Hs3}}V_{H}^{2}+\Lambda'_{13}V_{\Phi_{1}}^{2}+\Lambda'_{23}V_{\Phi_{2}}^{2}+\Delta_{13}V_{\phi_{1}}^{2}+\Delta_{23}V_{\phi_{2}}^{2}+2\lambda_{\text{s3}}V_{\phi_{3}}^{2}+\Delta_{3X}V_{\phi_{X}}^{2}+3\beta_{3X}V_{\phi_{3}}V_{\phi_{X}}\right.\nonumber \\
 &  & \left.-2\mu_{\text{s3}}^{2}\right)+\beta_{13}V_{\Phi_{1}}V_{\Phi_{2}}V_{\phi_{1}};\\
0 & = & V_{\phi_{X}}\left(\Lambda_{\text{Hsx}}V_{H}^{2}+\Lambda'_{1X}V_{\Phi_{1}}^{2}+\Lambda'_{2X}V_{\Phi_{2}}^{2}+\Delta_{1X}V_{\phi_{1}}^{2}+\Delta_{2X}V_{\phi_{2}}^{2}+2\lambda_{\text{sx}}V_{\phi_{X}}^{2}-2\kappa_{\text{H2X}}V_{H}V_{\Phi_{2}}-2\mu_{\text{sx}}^{2}\right)\nonumber \\
 &  & -\sqrt{2}\kappa_{\text{H1X}}V_{H}V_{\Phi_{1}}+\beta_{3X}V_{\phi_{3}}^{3}+\Delta_{3X}V_{\phi_{3}}^{2}V_{\phi_{X}};\label{vin2}
\end{eqnarray}
In the Eqs.\,(\ref{vin1})-(\ref{vin2}) above, $V_{H},\, V_{\Phi_{1}},\, V_{\Phi_{2}},\, V_{\phi_{1}},\, V_{\phi_{2}},\, V_{\phi_{3}},\, V_{\phi_{X}}$
are the VEVs of $H_{R}^{0},\,\Phi_{1R}^{0},\,\Phi_{2R}^{0},\,\phi_{1R},\,\phi_{2R},\,\phi_{3R}\,,\phi_{XR}$,
respectively.

\end{document}